\documentclass[12pt,english,journal,draftclsnofoot,onecolumn]{IEEEtran}
\usepackage[T1]{fontenc}
\usepackage[latin9]{inputenc}
\usepackage{float}
\usepackage{amsthm}
\usepackage{amsmath}
\usepackage{amssymb}
\usepackage{graphicx}
\usepackage{esint}
\usepackage[table]{xcolor}

\makeatletter
\theoremstyle{plain}

\usepackage{amsthm}

\theoremstyle{plain}

\IEEEoverridecommandlockouts
\@ifundefined{definecolor}{\usepackage{color}}{}
\usepackage{algorithmic}\usepackage{algorithm}

\usepackage{babel}
\providecommand{\theoremname}{Theorem}
\addto\captionsenglish{}

\usepackage{filecontents}

\begin{document}
\linespread{1}

\title{Next Generation M2M Cellular Networks: Challenges and Practical Considerations}

\author{{\normalsize {Abdelmohsen Ali, Walaa Hamouda, and Murat Uysal}} \thanks{%
A. Ali and W. Hamouda with the Dept. of Electrical and Computer Engineering, Concordia University, Montreal, Quebec, H3G 1M8, Canada, and M. Uysal with the Dept. of Electrical and Electronics Engineering, Ozyegin University, Istanbul, Turkey, 34794, e-mail:\{(ali\_abde,hamouda)@ece.concordia.ca\}, murat.uysal@ozyegin.edu.tr. }}

\maketitle
\linespread{1.667}

\thispagestyle{empty}
\begin{abstract}

In this article, we present the major challenges of future machine-to-machine (M2M) cellular networks such as spectrum scarcity problem, support for low-power, low-cost, and numerous number of devices. As being an integral part of the future Internet-of-Things (IoT), the true vision of M2M communications cannot be reached with conventional solutions that are typically cost inefficient. Cognitive radio concept has emerged to significantly tackle the spectrum under-utilization or scarcity problem. Heterogeneous network model is another alternative to relax the number of covered users. To this extent, we present a complete fundamental understanding and engineering knowledge of cognitive radios, heterogeneous network model, and power and cost challenges in the context of future M2M cellular networks.

\end{abstract}


\section{Introduction}

The Internet technology has undergone enormous changes since its early stages and it has become an important communication infrastructure targeting anywhere, anytime connectivity. Historically, human-to-human (H2H) communication, mainly voice communication, has been the center of importance. Therefore, the current network protocols and infrastructure are  optimized for human-oriented traffic characteristics. Lately, an entirely different paradigm of communication has emerged with the inclusion of "machines" in the communications landscape. In that sense, machines/devices that are typically wireless such as sensors, actuators, and smart meters are able to communicate with each other exchanging information and data without human intervention. Since the number of connected devices/machines is expected to surpass the human-centric communication devices by tenfold, machine-to-machine (M2M) communication is expected to be a key element in future networks ~\cite{ref:M2MIntr}. With the introduction of M2M, the next generation Internet or the Internet-of-Things (IoT) has to offer the facilities to connect different objects together whether they belong to humans or not.

\begin{figure}[H]
\centering
\includegraphics[scale=0.7]{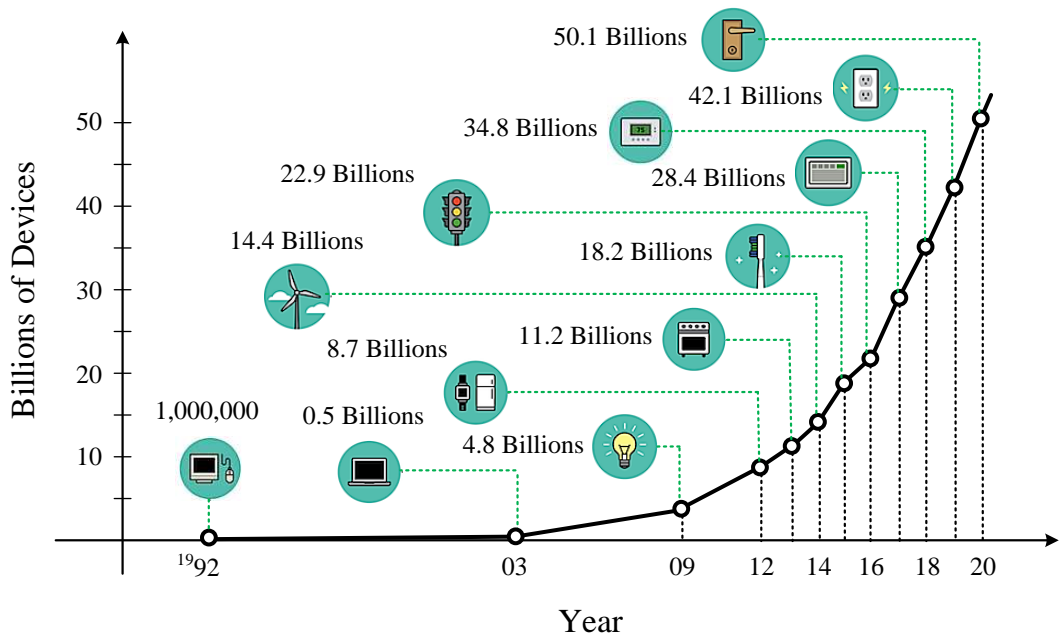}
\caption{Expected number of connected devices to the Internet. This chart is obtained from recent reports developed by both Cisco and Ericsson. The reports discuss the expected growth in the number of connected devices by 2020 due to the introduction of the M2M market.}
\label{fig:IoTNumDevices}
\end{figure}

The ultimate objective of M2M communications is to construct comprehensive connections among all machines distributed over an extensive coverage area. Recent reports show that the projected number of connected machines/devices to the IoT will reach approximately 50 billions by 2020 (Fig.~\ref{fig:IoTNumDevices}). This massive introduction of communicating machines has to be planned for and accommodated with applications requiring wide range of requirements and characteristics such as mobility support, reliability, coverage, required data rate, power consumption, hardware complexity, and device cost. Other planning and design issues for M2M communications include the future network architecture, the massive growth in the number of users, and the various device requirements that enable the concept of IoT. In terms of M2M, the future network has to provide machine requirements as power and cost are critical aspects of M2M devices. For instance, a set-and-forget type of application in M2M devices such as smart meters require very long battery life where the device has to operate in an ultra low-power mode. Moreover, the future network should allow for low complex and low data rate communication technologies which provide low cost devices that encourage the large scale of the IoT. The network architecture, therefore, needs to be flexible enough to provide these requirements and more. In this regard, a considerable amount of research has been directed towards available network technologies such as Zigbee (IEEE 802.15.4) or WiFi (IEEE 802.11b) by interconnecting devices in a form of large heterogeneous network~\cite{ref:hetroNetwork}. Furthermore, solutions for the heterogeneous network architecture (connections, routing, congestion control, energy-efficient transmission, etc.) have been presented to suit the new requirements of M2M communications. However, it is still not clear whether these sophisticated solutions can be applied to M2M communications due to constraints on the hardware complexity.

With the large coverage and flexible data rates offered by cellular systems, research efforts from industry have recently been focused on optimizing the existing cellular networks considering M2M specifications. Among other solutions, scenarios defined by the 3rd Generation Partnership Project (3GPP) standardization body emerge as the most promising solutions to enable wireless infrastructure of M2M communications~\cite{ref:3GPPInt}. In this front, a special category that supports M2M features has been incorporated by the 3GPP to Long-Term-Evolution (LTE) specifications. Due to the M2M communication challenges and the wide range of supported device specifications, developing the features for M2M communication, also refers to machine-type-communication (MTC) in the context of LTE, started as early as release 10 (R10) for the advanced LTE standard. This continued to future releases including release 13 (R13) that is currently developed and expected to be released in 2016. For these reasons, in this article, we will focus on the cellular MTC architecture based on the LTE technology as a key enabler with wide range of MTC support.

Due to the radical change in the number of users, the network has to carefully utilize the available resources in order to maintain reasonable quality-of-service (QoS). Generally, one of the most important resources in wireless communications is the frequency spectrum. To support larger number of connected devices in the future IoT, it is likely to add more degrees of freedom represented in more operating frequency bands. However, the frequency spectrum is currently scarce and requiring additional frequency resources makes the problem of supporting this massive number of devices even harder to solve. In fact, this issue is extremely important especially for the cellular architecture since the spectrum scarcity problem directly influences the reliability and the QoS offered by the network. To overcome this problem, \emph{small cell design}, \emph{interconnecting the cellular network to other wireless networks}, and \emph{cognitive radio (CR) support} are three promising solutions.

In this article, we address the issues that facilitate the existence of cellular MTC including the network architecture, the spectrum scarcity problem, and the device requirements. We review different approaches, including \emph{small cell design}, \emph{interconnecting the cellular network to other wireless networks}, and \emph{cognitive radio (CR) support}, based on research efforts and industrial technologies to tackle these issues. Furthermore, we provide a comparison of the potential solutions and the challenges and open issues that require future work to allow for practical development of each solution. The article is organized as follows. We provide an introduction to cellular MTC as well as the technological scenario of M2M communications based on the available standards. In the context of MTC, a description to the spectrum scarcity problem is discussed. This is followed by exposing the cognitive radio solution to nail this problem. We further present the cellular heterogeneous network concept. Then, important open issues and future directions are discussed. Finally, we draw our conclusions.


\section{Machine-Type-Communication in LTE Technology}
\label{sec:CellularMTC}

\begin{figure}[H]
\centering
\includegraphics[scale=0.46]{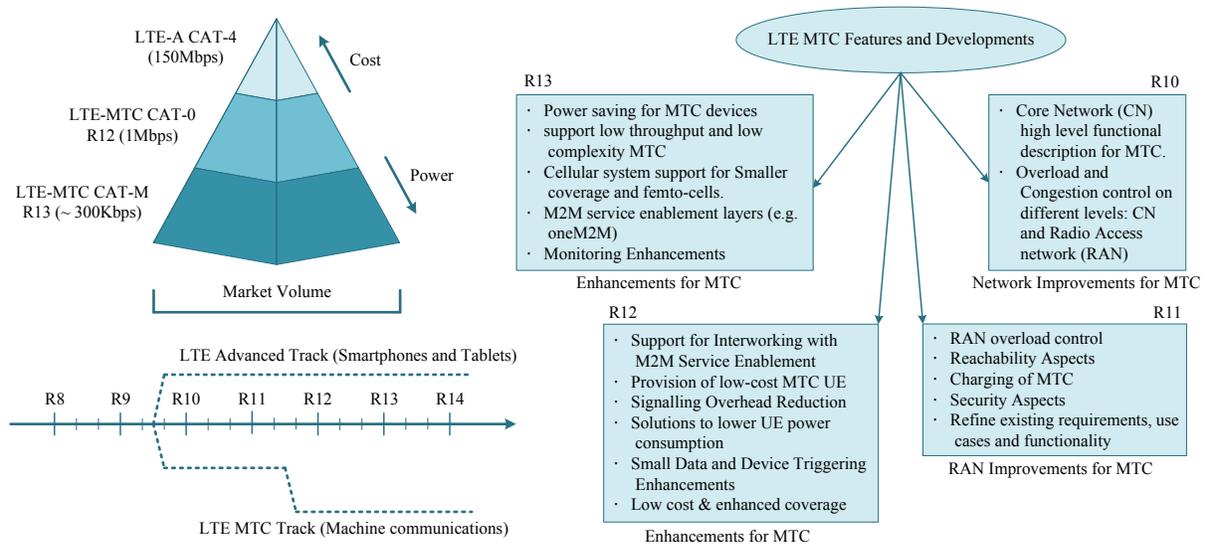}
\caption{MTC in 3GPP LTE Networks: Releases and Features.}
\label{fig:LTEMTC}
\end{figure}

Current M2M markets are highly fragmented and most vertical M2M solutions have been designed independently and separately for each application, which inevitably impacts large-scale M2M deployment~\cite{ref:M2Mbook}. However, when it comes to standardizations, the global coverage, the cellular network stability and maturity, together with the speed offered by recent cellular networks (LTE rates up to 150 Mbps for mobile objects), render wireless cellular technologies as the best candidate for the implementation of secure and reliable business critical M2M services. Several working groups in radio-access-networks (RAN) contribute very actively to the work on MTC-related optimization for 3GPP LTE networks. From day one, the support for MTC was one of the major concerns for the 3GPP and the development for a robust MTC design was divided across different releases~\cite{ref:M2MReleases}. Fig.~\ref{fig:LTEMTC} shows the development steps and features for MTC in different releases. Since LTE has the ability to support high performance, high throughput devices, the objective was to develop high volume, low cost, low complexity, and low throughput user-equipment (UE) LTE-based MTC devices.

From the history of MTC/LTE development, the first generation of a complete feature MTC device has emerged in R12. In this release, the 3GPP committee has defined a new profile referred to as category 0 or CAT-0 for low-cost MTC operation. Also a full coverage improvement is guaranteed for all LTE duplex modes. On the other hand, R13 is a future release for LTE-A in which MTC has the main weight of contribution. Its main goal is to further enhance the MTC LTE-based UE beyond R12. The main objectives for the MTC improvements are, (a) Supporting ultra low-power, low-cost, and narrow-band UE, (b) Enhancing the monitoring of service quality, and (c) Cooperation with other service delivery platforms represented in only the oneM2M~\cite{ref:oneM2MIntro}. Recall that the main objective of oneM2M is to minimize M2M service layer standards market fragmentation by consolidating currently isolated M2M service layer standards activities and jointly developing global specifications. In fact, seven of the worlds leading standards bodies, including for example the European Telecommunications Standards Institute (ETSI) and the Association of Radio Industries and Businesses (ARIB), have come together to create oneM2M. Although this solution considers some test cases for predefined devices such as smart metering, smart grids, eHealth, and automotive applications, not much attention has been taken on the underlying connectivity layer since oneM2M leverages on current and future technologies such as LTE networks.


\section{Small Cell Versus Heterogeneous Network Model}
\label{sec:SmallCellDeployment}

The next generation cellular MTC network has to efficiently interconnect several billions of wireless machines to support IoT. The traditional method to support these devices is to employ a well-designed M2M technology over a small cell structured system. In this case, cellular network providers need to deploy several thousands of base stations (eNodeB in LTE context) each with a smaller cell radius rather than full-power transmitters with large cells. Of course, this solution is cost-inefficient. Moreover, with such large number of small cells, co-channel interference is a limiting factor and complex designs are needed to maintain the required QoS. Another major drawback of this approach is the significant traffic increase due to signalling congestion and network management burden.

Although the "heterogeneous network" model is not currently recommended for MTC due to the limited capabilities of the machines, research efforts~\cite{ref:hetroWiFi} have been invested to support the idea of utilizing the cellular network itself as a small type of a heterogeneous network. The concept is that, in many applications, machines can be clustered geographically where the members of each cluster can be interconnected together through certain technology. To reduce the number of machines connected to the cellular network, each cluster would select a representative, a cluster head, to connect with the cellular network. Inside the cluster, the cellular network is transparent to all machines and only the cluster head will be responsible for relaying the aggregate traffic of the entire cluster. For example, if all machines have WiFi interfaces, then WiFi technology can be utilized to interconnect cluster members. In that sense, the cluster head will be communicating over its WiFi interface inside the cluster while using the LTE interface, for example, to connect to the cellular network (Fig.~\ref{fig:CombinedHetroAndH2HM2M}). In this model, the cellular network has offloaded part of its traffic to the individual clusters and therefore, reduces the effective number of covered users. The main benefit of this approach is the relaxation on congestion that would result if no clusters are formed.


\section{Cognitive Cellular M2M Networks}
\label{sec:CognitiveCellularM2MNetworks}

\begin{figure}[H]
\centering
\includegraphics[scale=0.47]{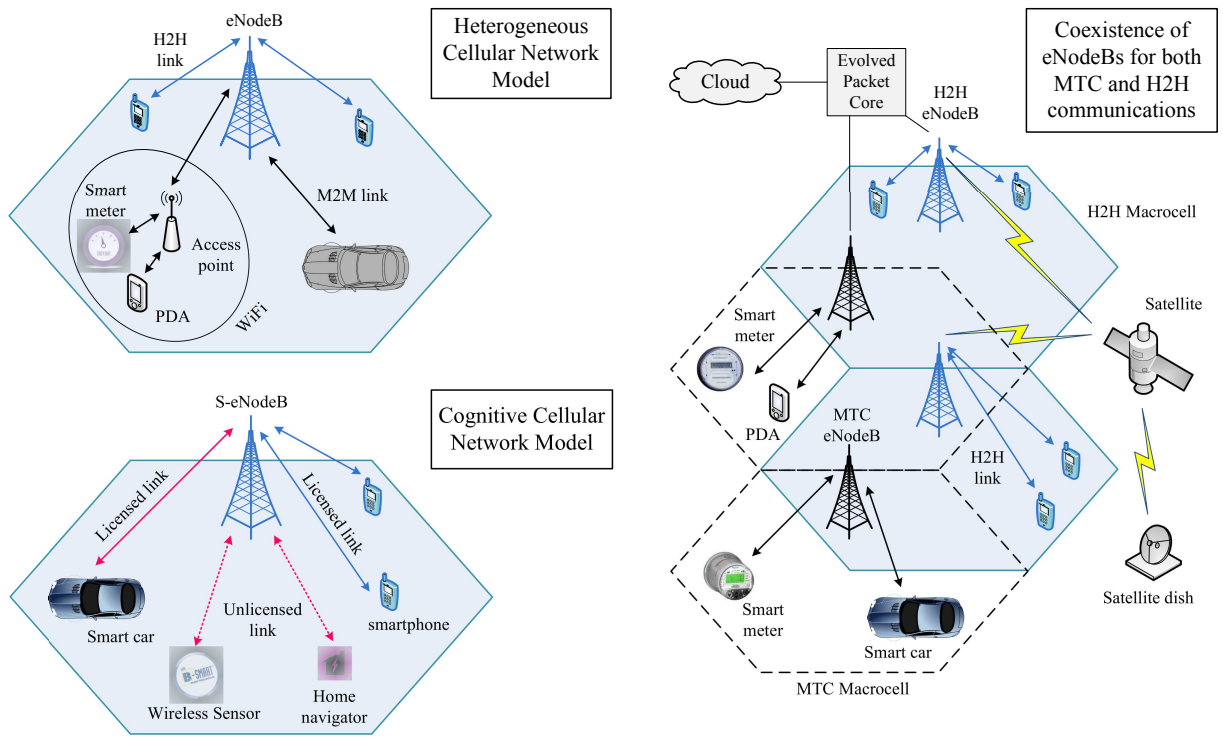}
\caption{Various network models to interconnect numerous number of machines to IoT.}
\label{fig:CombinedHetroAndH2HM2M}
\end{figure}

The idea of cognitive radios was originally proposed to offer more efficient utilization for the RF spectrum~\cite{ref:CRnet}. In this context, there are two approaches to apply the CR concept in cellular M2M networks. The first approach~\cite{ref:coexH2HM2M} assumes that there can be two types of eNodeB stations, one for typical UEs and other for MTC UEs coexisting with each other (Fig.~\ref{fig:CombinedHetroAndH2HM2M}), to relax signalling congestion and management burden. In this case, M2M devices seek to opportunistically use the spectrum when the H2H devices are idle. Therefore, M2M and H2H devices are not allowed to simultaneously operate over H2H links. This can be done through coordination between the corresponding eNodeB stations. Once a radio resource is occupied by M2M communications, this radio resource is regarded as suffering from server interference and will not be utilized by H2H communication. Even though this approach is simple to apply, it can degrade the QoS of H2H applications especially when the number of MTC devices is very large.

To overcome the aforementioned problems, we propose a second approach which supports unlicensed bands in addition to existing licensed bands. Here, it is assumed that the network will sense unlicensed bands to find extra vacant bands. If complexity permits, more than one unlicensed band per cell can be utilized by a Smart-eNodeB (S-eNodeB), a coined term to differentiate between the traditional eNodeB and the proposed one, to further increase the number of devices (Fig.~\ref{fig:CombinedHetroAndH2HM2M}). Indeed, this solution leverages on the huge amount of free spectrum available around the 5GHz and the TV white space. However, current radio access standards such as IEEE 802.22 and IEEE 802.11af already allow the use of this free unlicensed spectrum. Therefore, spectrum sensing and monitoring is a must. This can be implemented by introducing a new layer for spectrum management to support cognition over the unlicensed bands. That is, the S-eNodeB should be capable of (a) sensing the spectrum, (b) gathering information about the available suitable bands, (c) taking decisions on the conditions of these bands, (d) informing the neighboring S-eNodeB's about the allocated unlicensed band, (e) monitoring the allocated unlicensed band, and (f) always providing an alternative band.

\begin{figure}[H]
\centering
\includegraphics[scale=0.59]{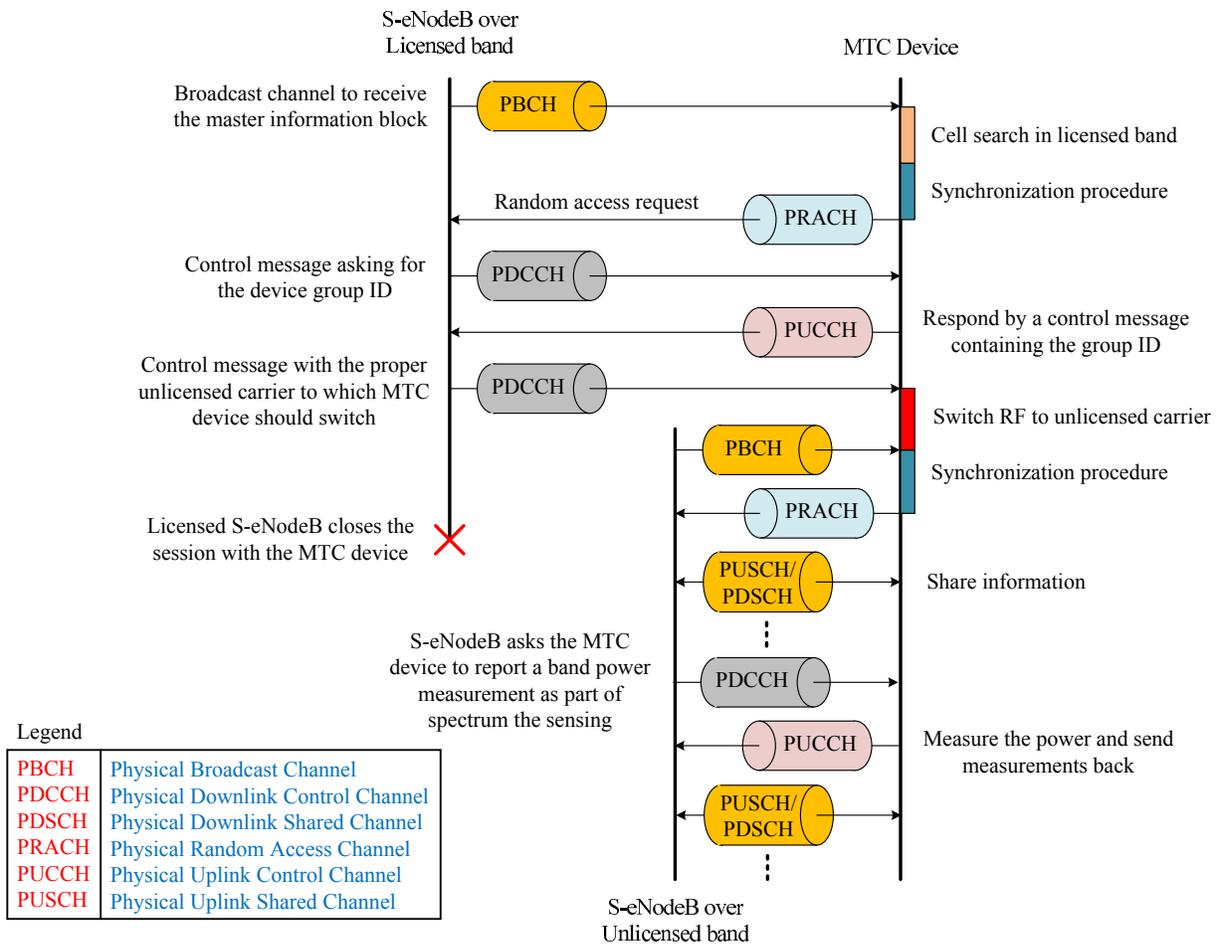}
\caption{Handshake messaging for MTC device over cognitive cellular network.}
\label{fig:TimingDiagram}
\end{figure}

If the S-eNodeB handles multiple unlicensed bands, then it should classify the machines based on their performance tolerance so that a machine is switched to the proper unlicensed band that meets its requirement. Of course, this assumes that the machine would have a group ID to declare its needs which in turn has to be shared with the S-eNodeB during call setup. To clarify how machines and S-eNodeB can work in this scenario, a detailed call procedure is demonstrated to show how a machine can access the unlicensed band. Once the machine is switched on, it goes to the calibration process in which the RF front-end adjusts or even estimates the IQ mismatch parameters. The following procedure is shown in Fig.~\ref{fig:TimingDiagram} and is discussed below.

\begin{itemize}
\item The machine would start the usual frequency scanning over the licensed LTE carriers. Once it locates a strong serving cell, a synchronization procedure is followed so that the machine is locked to the base station. It further decodes the master information block to recognize the cell specification.
\item The machine sends a random access request to connect to the cell. The Smart eNodeB then requests the group ID which will be sent over the uplink control channel.
\item The S-eNodeB will request the machine to switch to another carrier in the unlicensed band. Full information about the carrier such as modulation, coding, and relative timing to the licensed carrier are also sent to the machine. Afterwards, the S-eNodeB assumes that the session is complete and the machine has been configured.
\item The machine will then switch its RF to the desired carrier and enter the synchronization mode to lock itself to the S-eNodeB at the unlicensed carrier.
\item The machine defines itself one more time by sending a random access request over this carrier. If it is permitted, the machine can exchange data with the S-eNodeB over the physical uplink and downlink shared channels.
\item The S-eNodeB can interrupt the machine by scheduling a measurement gap in which the machine measures and reports the power of a certain carrier in the unlicensed or licensed bands.
\item The unlicensed carrier can be dynamically changed based on the collected measurements at the S-eNodeB. In this case, machines have to be informed about the new carrier and its settings.
\end{itemize}


\section{Ultra Low-Power and Low-Cost Networks}
\label{sec:UltraLowPowerLowCostNetworks}

To save battery life, low-power design is always desired for wireless communication systems. However, power reduction is not an easy task as it is related to the system reliability, rate of data exchange, and the radio chip design and implementation constraints. When the communication link is unreliable, higher layers translate this into retransmissions which results in longer active times and hence, high power consumption. Similarly, if the system continuously exchange data then it will consume more power. Based on a case study for ZigBee~\cite{ref:M2Mbook}, it is shown that if the radio is switched-on all the time, it will deplete a typical AA battery within a week. However, turning the radio duty cycle to 25\% extends the lifetime to about a month. Turning it further down to 1\% yields years of lifetime. Therefore, low power can be achieved through a reliable communication link with small duty cycle. In LTE-advanced, the concept of discontinuous reception (DRX) cycles is applied where the eNodeB schedules a silent period (DRX cycle) to encourage the device to switch-off the radio chip so that low duty cycle is achieved. To support ultra low-power design in recent releases, a long DRX cycle mode has been employed (with a maximum period of 2.56sec in R12) to further reduce the duty cycle.

Another important aspect of future MTC devices is their low-cost design which is typically provisioned by reducing the complexity of the system while providing the same coverage. The communication system architecture usually involves a general processor to run the software, a memory to hold both instructions and data, and a physical-layer modem to handle the communication protocol. As expected, most of the complexity reduction comes from the physical-layer modem features along with a small portion of data memory reduction. Therefore, a low-cost design is typically related to a feature reduction while the coverage is carefully kept unchanged. For specific applications, low data rates and/or low latency are acceptable. In this case, the modem features can be relaxed to target low-cost design. In recent LTE releases, special category has been defined to support MTC for low data rates which leads to complexity reduction. In LTE-R12, this category supports only one receive antenna and a maximum data rate of 1Mbps. However, those features will be further reduced in LTE-R13 with the expected maximum data rate being 300Kbps and only one operating bandwidth of 1.4MHz.


\section{Challenges, Open Issues, and Future Directions}
\label{sec:OpenIssuesFutureDirections}

\subsection{Heterogeneous Networks}

When a number of machines is able to form clusters, the cellular network becomes lightly loaded. This conclusion has been investigated by many researchers and even practically demonstrated on WiFi as the internal technology inside the cluster. However, it is hard to judge if the machines can really form clustering or not. In fact, clusters are formed only if the WiFi connectivity between cluster members is acceptable (data rates are higher than the LTE load generated in the cluster). Also, clustering allows machines to enjoy seamless connectivity to the cellular system while spending more time on a secondary, WiFi-based interface, which is less power consuming than LTE. On the other hand, shifting the responsibility of the aggregate traffic from all cluster members to the cluster head can be challenging especially if the link from the cluster head to the eNodeB is poor. Since the architecture assumes a centralized control at the head node, it is expected that the full cluster will fail. Therefore, more research effort is required to investigate the possibility of dynamically selecting the head node based on the channel quality with the cellular system. One challenge with this solution is to select the optimum period after which a rescheduling has to be done.

\subsection{Cognitive Radio Network}

As discussed earlier, spectrum sensing and monitoring are essential to utilize the cognitive radio concept in which some of the machines operate over an unlicensed band. However, there are many challenges to address this problem.

\begin{figure}[H]
\centering
\includegraphics[scale=0.7]{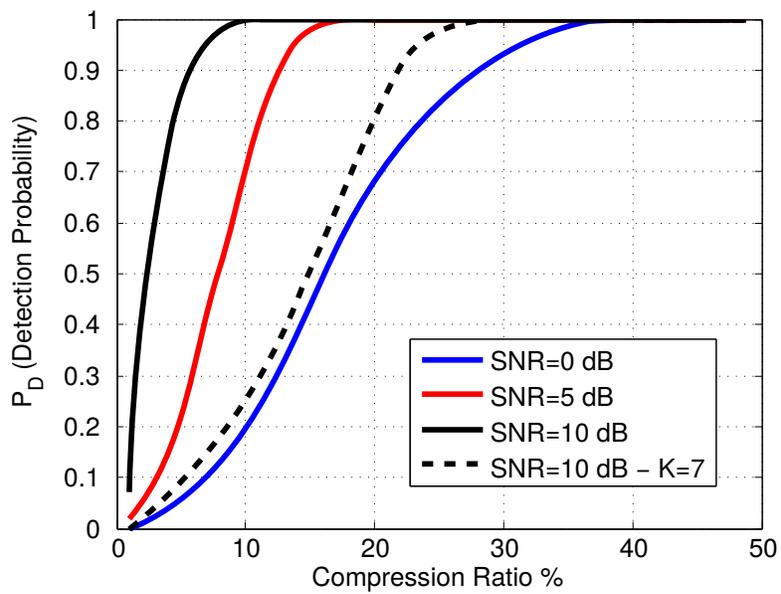}
\caption{Effect of compression ratio (i.e., ratio of the non-uniform sampling rate to the conventional Nyquist rate) on the detection performance when the false alarm rate is 1\%. 16 contiguous non-overlapped bands are investigated where each has a bandwidth of 1MHz. Only four active bands are considered, therefore the sparsity level $K=4$ out of the available 16}
\label{fig:CompressedSensing}
\end{figure}

\begin{figure}[H]
\centering
\includegraphics[scale=0.5]{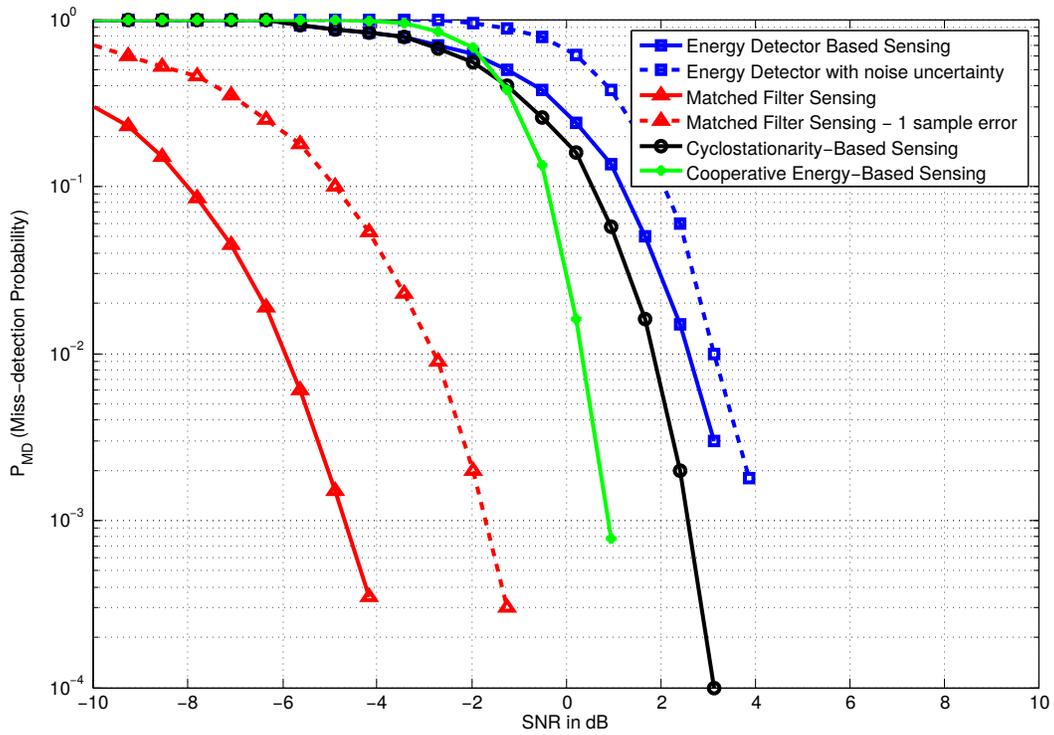}
\caption{These curves are plotted for a false alarm rate of 1\%. The window size for the energy detector is the same as the matched filter length. Both agree with the cyclostationary detector period which is 32 samples. For the cooperative sensing, hard decision is used with K-out-of-N rule where K=5 users and N=10 users. The noise uncertainty error is $\pm$0.5 dBs for the energy detection case.}
\label{fig:CompareNarrowBand}
\end{figure}

\begin{itemize}
\item \emph{Spectrum sensing techniques}: The sensing can be either centralized at the S-eNodeB or done in cooperation with the machines. Better performance is expected from the latter case since more spatial diversity is utilized. Generally, cooperation is achieved by sending either local decisions~\cite{ref:cooperSensingCompressed}, which can be either hard or soft decision, or by sending the useful portion of the received data set. Processing power of the machines limits the first approach while high traffic over the control channel is the main challenge for the second approach. Moreover, the link between the machine and the S-eNodeB is not ideal and the sensing decisions/data can be received incorrectly which may alter the sensing accuracy at the S-eNodeB.

\item \emph{Wideband sensing methodology}: During the initial sensing stage, a very wideband (around 1GHz) has to be assessed to locate a suitable vacant band. This can be implemented by scanning different bands one after another and measuring the in-band power. This technique is simple but it requires time and power to find a suitable band. Another alternative is to examine the power spectral density of the entire wideband at once. Since this method requires high speed analogue-to-digital conversion, compressive sensing (CS)~\cite{ref:cooperSensingCompressed} is a promising technique to obtain the power spectral density of the wideband spectrum while sampling at rates lower than the Nyquist rate. The concept is to capture few measurements of the \emph{sparse} spectrum. The wideband spectrum is related to those raw measurements by a linear under-determined system of equations. Optimization techniques can be employed to solve this set of equations in order to find the best solution that satisfies the original assumption for the spectrum which is being \emph{sparse}. Fig.~\ref{fig:CompressedSensing} shows the detection performance as a function of the ratio of non-uniform sampling frequency to the typical Nyquist rate. It is clear that CS is able to detect the spectrum occupancy by a ratio of 1/10 of the Nyquist rate at high signal-to-noise ratio (SNR). Although CS is very promising in this context, many challenges exist due to the current algorithmic complexity as well as the basic assumptions. For example, the spectrum is dynamically loaded and the \emph{sparse} assumption may not be valid which results in performance degradation ($K=4,7$ cases in Fig.~\ref{fig:CompressedSensing}). Cooperation may be utilized to enhance the accuracy, however finding a high-performance low-complex/low-data rate cooperative sensing technique is not a trivial task. In that direction, more research efforts are needed to develop efficient algorithms to render CS possible with reasonable complexity, especially for MTC where complexity is a real challenge.

\item \emph{Narrowband sensing techniques}: A signal processing algorithm is needed to decide on the activities within each of the wideband slices (vacant or not). Conventional algorithms/detect- ors~\cite{ref:spectrumsensingAlg} include the energy detector, the cyclostationary detector, and the matched-filter detector. In all cases, a decision statistic is computed and compared to a threshold to decide whether a specific band is occupied or not. Complexity, performance, and prior information about the signal to be detected are the main metrics to judge the quality of the detector. Among those detectors, the energy detector is known to be the only simple non-coherent detector. From performance perspective, the matched-filter is known to be the optimal detector. However, it requires full knowledge of the detected signal.

The Cyclostationary detector can be used only if the signal possesses the cyclostationarity property where its statistics, mean and autocorrelation, are periodic with some known period. Therefore, it requires partial information about the detected signal which is typically the period of cyclostationarity. Fig.~\ref{fig:CompareNarrowBand} shows the probability of miss-detection for various narrowband sensing techniques against SNR. Effects of timing errors, noise uncertainty, and hard decision cooperative sensing have been included. The performance results show that: (a) Any uncertainty of the noise level will significantly alter the performance of the energy detector, (b) Matched-filter detection is very sensitive to timing errors, and (c) Cooperation involves high diversity gain. However, these results assume an ideal channel (no noise and no fading) between the machines and the S-eNodeB. The conclusion is that, improvements and/or new sensing techniques are needed to provide less-complex, non-coherent, and robust practical algorithms.
\end{itemize}

\subsection{Low-Power Low-Cost Networks}

Although longer DRX cycle significantly reduces the power, it also introduces some challenges to the system design. Since the radio chip will be off during the DRX cycle, the device/UE has no way to synchronize itself to the eNodeB. Therefore, the typical behaviour for the device/UE would be to wake-up as early as required to quickly resynchronize itself to the eNodeB before receiving further packets. One of the issues is to determine the best wake-up time so that the synchronization performance is met and no additional power is lost. Another issue is related to the cooperative sensing architecture, if applicable, where the device/UE will not be able to sense or monitor any band while it is in a deep sleep mode. The band can suffer from high interference levels caused by other networks that attempt to access the same band. Finally, the power can be minimized by properly designing power domains in the hardware to decide which module is not needed to be switched-off.

A low-cost design always comes at the account of system performance with less features provided. For instance, reducing the number of receive antennas from two to one would reduce the spatial diversity of the modem. Therefore, advanced signal processing algorithms for synchronization, cell detection, and decoding will need to be revised to guarantee the same performance with less diversity gain. Indeed, reducing the cost is not only related to the required features from the network, but it also depends on the hardware design process and underlaying technology. For example, optimizing the internal word sizes of the various hardware modules inside the modem will result in a low gate count and low power consumption. However, the optimization algorithms that can achieve this are not unique as signal statistics across various modules are system dependent.


\section{Conclusion}
\label{sec:conclusion}
We presented the challenges that are expected from the next generation MTC network as an integral part of the future IoT. It is argued that cognitive radio concept is a possible solution from the cost and performance perspectives. However, there are more practical challenges that need efforts from researchers. The application of the heterogeneous network concept was investigated where cellular MTC networks can utilize other networks such as WiFi to reduce the number of directly connected machines/users. Future standards are encouraged to provide both options (i.e. the cognition concept and the heterogeneous network model). Finally, a design of low-power low-cost machine is discussed. However, there are important design challenges to make it possible. For example, an extended DRX cycle is a valid option to significantly reduce the power budget. No matter how, a feasibility study is scheduled in Release 13 to provide a solid solution in which extended DRX cycles implementation challenges can be overcome, if possible. The trade-off between cost, feasibility, and performance have also been discussed.

\linespread{1.6}
\bibliography{MyReferences}

\begin{thebibliography}{10}
\providecommand{\url}[1]{#1}
\csname url@samestyle\endcsname
\providecommand{\newblock}{\relax}
\providecommand{\bibinfo}[2]{#2}
\providecommand{\BIBentrySTDinterwordspacing}{\spaceskip=0pt\relax}
\providecommand{\BIBentryALTinterwordstretchfactor}{4}
\providecommand{\BIBentryALTinterwordspacing}{\spaceskip=\fontdimen2\font plus
\BIBentryALTinterwordstretchfactor\fontdimen3\font minus
  \fontdimen4\font\relax}
\providecommand{\BIBforeignlanguage}[2]{{%
\expandafter\ifx\csname l@#1\endcsname\relax
\typeout{** WARNING: IEEEtran.bst: No hyphenation pattern has been}%
\typeout{** loaded for the language `#1'. Using the pattern for}%
\typeout{** the default language instead.}%
\else
\language=\csname l@#1\endcsname
\fi
#2}}
\providecommand{\BIBdecl}{\relax}
\BIBdecl

\bibitem{ref:M2MIntr}
J.~Wan, D.~Li, C.~Zou, and K.~Zhou, ``{M2M} {C}ommunications for {S}mart
  {C}ity: {A}n {E}vent-{B}ased {A}rchitecture,'' in \emph{IEEE International
  Conference on Computer and Information Technology (CIT)}, Oct 2012, pp.
  895--900.

\bibitem{ref:hetroNetwork}
A.~Damnjanovic, J.~Montojo, Y.~Wei, T.~Ji, T.~Luo, M.~Vajapeyam, T.~Yoo,
  O.~Song, and D.~Malladi, ``A survey on {3GPP} heterogeneous networks,''
  \emph{IEEE Wireless Communications}, vol.~18, no.~3, pp. 10--21, June 2011.

\bibitem{ref:3GPPInt}
S.~Y. Lien, K.~C. Chen, and Y.~Lin, ``Toward ubiquitous massive accesses in
  {3GPP} machine-to-machine communications,'' \emph{IEEE Communications
  Magazine}, vol.~49, no.~4, pp. 66--74, April 2011.

\bibitem{ref:M2Mbook}
C.~A. Haro and M.~Dohler, \emph{{M}achine-to-{M}achine ({M2M})
  {C}ommunications: {A}rchitecture, {P}erformance and {A}pplications ({G}oogle
  eBook)}.\hskip 1em plus 0.5em minus 0.4em\relax Elsevier, Dec 2014.

\bibitem{ref:M2MReleases}
3GPP, ``Standarization of {M}achine-type {C}ommunications,'' 3rd Generation
  Partnership Project, Tech. Rep. V0.2.4, June 2014.

\bibitem{ref:oneM2MIntro}
J.~Swetina, G.~Lu, P.~Jacobs, F.~Ennesser, and J.~Song, ``Toward a standardized
  common {M2M} service layer platform: Introduction to {oneM2M},'' \emph{IEEE
  Wireless Communications}, vol.~21, no.~3, pp. 20--26, June 2014.

\bibitem{ref:hetroWiFi}
A.~Pyattaev, K.~Johnsson, S.~Andreev, and Y.~Koucheryavy, ``{3GPP} {LTE}
  traffic offloading onto {WiFi} {D}irect,'' in \emph{IEEE Wireless
  Communications and Networking Conference Workshops}, April 2013, pp.
  135--140.

\bibitem{ref:CRnet}
B.~Wang and K.~Liu, ``Advances in cognitive radio networks: {A} survey,''
  \emph{IEEE Journal of Selected Topics in Signal Processing}, vol.~5, no.~1,
  pp. 5--23, Feb 2011.

\bibitem{ref:coexH2HM2M}
S.~Y. Lien, S.~M. Cheng, S.~Y. Shih, and K.~C. Chen, ``{R}adio {R}esource
  {M}anagement for {QoS} {G}uarantees in {C}yber-{P}hysical {S}ystems,''
  \emph{IEEE Transactions on Parallel and Distributed Systems}, vol.~23, no.~9,
  pp. 1752--1761, Sept 2012.

\bibitem{ref:cooperSensingCompressed}
I.~F. Akyildiz, B.~F. Lo, and R.~Balakrishnan, ``Cooperative {S}pectrum
  {S}ensing in {C}ognitive {R}adio {N}etworks: {A} {S}urvey,'' \emph{Phys.
  Commun.}, vol.~4, no.~1, pp. 40--62, Mar. 2011.

\bibitem{ref:spectrumsensingAlg}
T.~Yucek and H.~Arslan, ``A survey of spectrum sensing algorithms for cognitive
  radio applications,'' \emph{IEEE Communications Surveys Tutorials}, vol.~11,
  no.~1, pp. 116--130, First 2009.

\end{thebibliography}
 \bibliographystyle{IEEEtran}
\linespread{1.6}

\pagebreak

\vspace*{\fill}
\vspace*{\fill}

\pagebreak

\vspace*{\fill}
\vspace*{\fill}

\pagebreak

\vspace*{\fill}
\vspace*{\fill}

\pagebreak

\vspace*{\fill}
\vspace*{\fill}

\pagebreak

\vspace*{\fill}
\vspace*{\fill}

\pagebreak

\vspace*{\fill}
\vspace*{\fill}

\end{document}